%% file: paper.tex
\documentclass[11pt,a4paper]{article}
\usepackage[margin=24mm]{geometry}
\usepackage{enumitem}
\usepackage{amsmath,amssymb,amsthm}
\usepackage{xcolor}
\usepackage{braket}
\usepackage{tikz}
\usepackage{titlesec}
\usepackage[style=numeric, sorting=none]{biblatex}
\bibliography{refs}
\definecolor{linkblue}{RGB}{25,63,98}
\usepackage[colorlinks=true,allcolors=linkblue]{hyperref}
\newcommand{\poly}{\operatorname{poly}}
\newtheorem{theorem}{Theorem}
\newtheorem{lemma}[theorem]{Lemma}
\newtheorem{corollary}[theorem]{Corollary}
\newtheorem{proposition}[theorem]{Proposition}
\titleformat*{\section}{\normalfont\large\bfseries}

\title{\makebox[\textwidth][c]{Implications of Sub-Exponential Counting
for Quantum Advantage}}
\author{Konstantinos Meichanetzidis$^{1a}$, Tuomas Laakkonen$^{2b}$,\\
Richie Yeung$^1$, Marcello Benedetti$^1$, Harry Buhrman$^1$\\
~\\
\emph{\small$^1$Quantinuum}, ~\emph{\small$^2$Massachusetts Institute of Technology}\\
{\small Email: \href{mailto:k.mei@quantinuum.com}{${}^a$\texttt{k.mei@quantinuum.com}}, \href{mailto:tsrl@mit.edu}{${}^b$\texttt{tsrl@mit.edu}}}}
\date{November 30, 2026}

\newcommand{\binarytosatstatement}{%
    Suppose that $T$ is a scalar tensor network with $t$ tensors, where each bond has dimension bounded by a constant, and each tensor has entries drawn from $\{0,
    1\}$ and degree bounded by a constant $2k$. Then there exists a sparse 3-CNF formula
    $\phi({\mathbf{x}})$ with $n = O(t)$ variables and $m = O(t)$ clauses,
    such that the number of satisfying solutions $|\{{\mathbf{x}} \mid
    \phi({\mathbf{x}}) = 1\}|= \mathrm{Contract}(T)$.%
}
\newtheorem*{restatedbinarytosat}{Lemma~\ref{thm:binarytosat} (restated)}

\newcommand{\nonbinarytobinarystatement}{%
    Let $k$ and $d$ be constants, and set $\zeta_d = e^{\frac{2\pi i}{2^d}}$. Suppose that $T$ is a
    {scalar} tensor network with $t$ tensors, where each bond has dimension two,
    and each tensor $T_i$ has degree bounded by $2k$. {Let $\mathbf{x}_i$ denote
    the indices of $T_i$. Its entries have the form}
    \[
        {T_i(\mathbf{x}_i)}
        =\gamma_i\sum_{j=0}^{{2^d-1}}
        {\beta^{(i)}_{j,\mathbf{x}_i}}\zeta_d^j,
    \]
    where ${\beta^{(i)}_{j,\mathbf{x}_i}} \in \{0, 1\}$ and $\gamma_i \in
    \mathbb{C}$ is a per-tensor scalar. Then there exists a set of $2^d = O(1)$ binary-valued tensor
    networks ${T^{(0)},\dots,T^{(2^d-1)}}$ and coefficients
    ${\alpha_0,\dots,\alpha_{2^d-1}} \in \mathbb{C}$ such that:
    $$ \mathrm{Contract}(T) =
    {\sum_{r=0}^{2^d-1}\alpha_r\cdot\mathrm{Contract}(T^{(r)})}$$
    Each ${T^{(r)}}$ can be calculated in $O(t)$ time and contains $O(t)$
    tensors with degree bounded by $2k + O(1)$, {bond dimensions in
    $\{2,2^d\}$,} and entries in $\{0, 1\}$, so that it satisfies the constraints of Lemma
    \ref{thm:binarytosat}.%
}
\newtheorem*{restatednonbinarytobinary}{Lemma~\ref{thm:nonbinarytobinary} (restated)}

\newcommand{\algebraicgatesstatement}{%
    Fix a finite set $\mathcal{G}$ of gates acting on a bounded number of qubits and a finite set
    $\mathcal{B}$ of single-qubit states, all with {complex entries}. For an $n$-qubit circuit $U$
    with $G$ gates from $\mathcal{G}$ and product states $\ket{\psi},\ket{\phi}$ with factors in
    $\mathcal{B}$, let $s$ be the number of distinct nonzero tensor entries in the construction of
    Lemma~\ref{thm:circuittononbinary}. Then the amplitude $\bra{\psi}U\ket{\phi}$ can be computed
    exactly using $O((n+G)^{s-1})$ \#3-SAT queries, each with $O(n+G)$ variables and clauses, and
    {polynomially many additional arithmetic operations, assuming exact arithmetic
    on the entries.} The polynomial degrees and constants may depend on these fixed sets.%
}
\newtheorem*{restatedalgebraicgates}{Proposition~\ref{thm:algebraicgates} (restated)}

\begin{document}
\maketitle
\begin{abstract} 

Recent preprints from OpenAI and Anthropic claim to give sub-exponential time algorithms for exact model counting. We explain how these bounds transfer to classical simulation of quantum circuits: for fixed gate sets, including Clifford+T, the resulting runtime is sub-exponential in the qubit count $n$ when the gate count satisfies $o(n\sqrt{\log n})$.
We comment on how popular quantum algorithms are excluded from this regime, and why direct SAT encodings of problems such as factoring yield better classical runtime than simulating the corresponding quantum algorithms, regardless of potential further improvements to model counting algorithms. Hence, we argue that the reported results improve particular simulation bounds without substantially changing the landscape of quantum advantage.
\end{abstract}

\section{Introduction}
\label{sec:introduction}

OpenAI and Anthropic recently released manuscripts \cite{OpenAI,Anthropic} claiming to disprove the Exponential Time Hypothesis (ETH). In fact, they both make the stronger claim that \#k-SAT can be solved in sub-exponential time (i.e. $2^{o(n)}$) with respect to the number of variables $n$, for every fixed value of $k$. These results are constructive, giving concrete algorithms for solving \#k-SAT which run in sub-exponential time. Since a wide variety of problems have reductions to k-SAT or \#k-SAT, this has far-reaching implications for fine-grained complexity. One such problem is the classical simulation of quantum circuits, which can be reduced to \#3-SAT, and thus benefits from the new algorithms. Therefore, it seems natural to ask whether these results have an impact on quantum computing, i.e.:

\begin{quote}
\centering
    \emph{Has quantum advantage become more difficult to achieve?}
\end{quote}

In short, no, the current landscape of quantum advantage probably has not been significantly altered. However, this statement is not entirely without caveats. In this note we will describe how \#3-SAT is connected to simulating quantum circuits, which circuits can now be simulated more efficiently, and whether this is likely to impact well-known quantum algorithms. We will mainly consider worst-case asymptotic time complexity here, since the algorithms of \cite{OpenAI,Anthropic} are unlikely to be practical, and remark briefly on heuristic methods in Section \ref{sec:discussion}.

\section{Classical simulation of quantum circuits}
\label{sec:reduction}

It has been known since Bernstein and Vazirani's \cite{BV} initial development of quantum complexity theory that, given access to an oracle for solving a \#P-complete problem such as \#3-SAT, a quantum computer can be simulated efficiently. Since then, improvements have been made to the proof, for example by Fortnow and Rogers \cite[Section 3]{FR}, and Dawson et al. \cite{Dawson}. Recently, this method has been implemented in practice \cite{MBL} using state-of-the-art \#SAT solvers, although it is not particularly competitive with alternative algorithms. 

There are several different senses in which a quantum computer may be classically simulated, e.g. calculating circuit amplitudes, sampling measurement outcomes, or estimating observables. Approximating an amplitude of a polynomial-sized quantum circuit to inverse-polynomial additive error is BQP-hard, and the corresponding decision problem is BQP-complete. Thus circuit amplitudes, which we consider now, provide a natural target for classical simulation algorithms.

An instructive lens through which to see these methods is \emph{scalar tensor networks} -- i.e. tensor networks where all indices are contracted. For a scalar tensor network $T$, we denote this contracted value as $\mathrm{Contract}(T)$. For arbitrary product states $\ket{\psi}$ and $\ket{\phi}$, the amplitudes $\bra{\psi}U\ket{\phi}$ of a quantum circuit can be encoded in the contracted value of a tensor network where each tensor corresponds to one quantum gate.  The construction is originally due to Markov and Shi \cite{MS}, and can be formalized as follows:

\begin{lemma}
    \label{thm:circuittononbinary}
    Let $\mathcal{G} = \{g_1, g_2, \dots, g_m\}$ be a set of quantum gates acting on at most $k$
    qubits whose unitaries have entries from a finite set $\mathcal{S} \subset \mathbb{C}$, and
    $\{\ket{\psi_i}\}_i$ be a set of single-qubit states whose state-vectors have entries from
    $\mathcal{S}$. For any $n$-qubit circuit $U$ with $G$ gates over $\mathcal{G}$, and any pair of
    $n$-qubit product states $\ket{\psi}, \ket{\phi}$ drawn from $\{\ket{\psi_i}\}_i$, there exists
    a scalar tensor network $T$ such that $\mathrm{Contract}(T) = \bra{\psi}U\ket{\phi}$. $T$
    contains $O(n + G)$ tensors, each with at most $2k$ indices (i.e. the degree) and entries lying
    in {$\mathcal{S}\cup\overline{\mathcal{S}}$, where
    $\overline{\mathcal{S}}=\{\overline{s}:s\in\mathcal{S}\}$}; all indices in $T$ have dimension
    two.
\end{lemma}

\begin{proof}
    See e.g. Markov and Shi \cite[Section 3]{MS}.
\end{proof}

\#SAT and other constraint satisfaction counting problems can also be naturally encoded as the contracted value of a scalar tensor network \cite{Kourtis2019} (i.e. consider the truth table of each clause as elements of a
tensor).
For our reduction, however, we need the converse direction: encode tensor network contractions to \#SAT. This can be done to some extent, depending on the entries of each tensor and their degree; that is why we introduced these parameters in the previous lemma. One result that is easy to see is the following, for binary-valued tensor networks:

\begin{lemma}
    \label{thm:binarytosat}
    \binarytosatstatement
\end{lemma}

\begin{proof} See Appendix~\ref{app:binarytosat}. \end{proof}

Unfortunately, all the interesting
gate sets for quantum computing require non-binary valued tensors -- the only unitaries with binary
entries are permutations, and quantum circuits composed of these represent classical reversible
logic, which is efficiently simulable when {the input is a computational basis
state}. One way to circumvent this is to write the contraction of a more general tensor network as a
weighted sum of contractions of {binary-valued} tensor networks:

\begin{lemma}
    \label{thm:nonbinarytobinary}
    \nonbinarytobinarystatement
\end{lemma}

\begin{proof} See Appendix~\ref{app:nonbinarytobinary}. An alternative proof using the ZH-calculus is given in \cite[Section~3]{LMW}. \end{proof}

This Lemma allows for useful gate sets to be encoded -- for example, Toffoli+H with $d = 1$ and $k = 3$, and Clifford+T with $d = 3$ and $k = 2$. By combining Lemmas \ref{thm:circuittononbinary}, \ref{thm:binarytosat}, and \ref{thm:nonbinarytobinary}, we can see how the time necessary to simulate a quantum circuit scales, given access to a \#3-SAT solver:

\begin{theorem}
    \label{thm:amplitudetosat}
    Let $k$ and $d$ be constants, and set $\zeta_d = e^{\frac{2\pi i}{2^d}}$. Define $\mathcal{G} =
    \{g_1, g_2, \dots, g_m\}$, a set of quantum gates acting on at most $k$ qubits, and
    $\{\ket{\psi_i}\}_i$ a set of single-qubit states, where the unitaries of $g_i$ and
    state-vectors of $\ket{\psi_i}$ have entries of the form $\gamma \sum_{j =
    0}^{{2^d-1}} \beta_{j}\zeta^j_d$, where $\beta_{j} \in \{0, 1\}$ and
    $\gamma \in \mathbb{C}$ is a scalar shared between all entries in the same unitary or
    state-vector.
    Suppose that we have access to a \#3-SAT solver which runs in time $T_{\#\mathrm{3-SAT}}(v)$ for sparse instances with $v$ variables and $O(v)$ clauses. For any $n$-qubit circuit $U$ with $G$ gates over $\mathcal{G}$, and any pair of $n$-qubit product states $\ket{\psi}, \ket{\phi}$ drawn from $\{\ket{\psi_i}\}_i$, the amplitude $\bra{\psi}U\ket{\phi}$ can be evaluated in time at most:
    $$ 2^{d}\cdot T_{\#\mathrm{3-SAT}}(O(n + G)) + O(2^d) $$
\end{theorem}

\begin{proof}
    Apply Lemma \ref{thm:circuittononbinary} to $U$, followed by {Lemma
    \ref{thm:nonbinarytobinary} and then Lemma \ref{thm:binarytosat}}. Solve the resulting \#3-SAT
    instances to obtain counts, and reconstruct the final amplitude from the counts and coefficients
    ${\alpha_r}$.
\end{proof}

{
{
Theorem~\ref{thm:amplitudetosat} can be used to approximately simulate other gate sets by
synthesizing their gates as Clifford+T circuits. For example, approximating each of $K$ gates 
to error $\varepsilon/K$ gives total error at most $\varepsilon$ and requires
$O(K\log(K/\varepsilon))$ Clifford+T gates. This overhead increases the size of the resulting
counting instances. Alternatively, for any fixed finite gate set with complex entries, we can avoid this synthesis overhead and
compute amplitudes exactly, if we can do exact arithmetic on the entries. The following proposition uses polynomially many counting queries, each of size linear in the original circuit size:
}

\begin{samepage}

\begin{proposition}
    \label{thm:algebraicgates}
    \algebraicgatesstatement
\end{proposition}

\begin{proof} See Appendix~\ref{app:algebraicgates}. \end{proof}

\end{samepage}
}

Rather than \#3-SAT, it is also possible to reduce quantum circuit simulation to counting the solutions of sparse polynomial equations over finite fields \cite{Dawson}. 
Refs. \cite{OpenAI,Anthropic} also provide an algorithm for this problem, and the recent paper \cite{Gheorgiu2026} discusses its consequences for quantum circuit simulation.

\section{Which circuits are now simulable in sub-exponential time?}
\label{sec:simulable}

The alternative algorithm for classical simulation of quantum circuits that we will compare to is state-vector simulation; given an $n$-qubit circuit with $G$ gates, this operates in time $O(2^n \cdot G)$. Crucially, $G$ is in the prefactor of the exponential, and \emph{not} the exponent -- hence, given an algorithm for \#3-SAT which runs in time exponential in the number of variables $v = O(n + G)$, Theorem \ref{thm:amplitudetosat} is never better than state-vector simulation when $G$ is super-linear in $n$.

However, given a sub-exponential algorithm, there exists a small window where the \#SAT-based method can win out, at least asymptotically. To analyze where this occurs, we need to know the particular sub-exponential scaling claimed by OpenAI and Anthropic. A special case of their results can be formulated as follows:

\begin{theorem}
    \label{thm:slopstatement}
    For a sparse 3-CNF instance with $v$ variables and at most $O(v)$ clauses, there is an algorithm which solves \#3-SAT in time:
    $$ T_{\#\mathrm{3-SAT}}(v) = \exp \left[O\left(\frac{v}{\sqrt{\log v}}\right)\right] $$
\end{theorem}

\begin{proof}
    See \cite[Theorem 1.1]{OpenAI} and \cite[Corollary 11]{Anthropic}.
\end{proof}

When reducing circuit simulation to \#3-SAT, a runtime bound that is subexponential in $v$ may fail to be subexponential in the qubit count $n$, even if $v$ is polynomial in $n$.
We therefore need to track how $v$ grows with both $n$ and the gate count $G$.
Applying this to Theorem \ref{thm:amplitudetosat} we arrive at the following result:

\begin{corollary}
    \label{thm:subexpsim}
    Let $U$ be an $n$-qubit quantum circuit with $G$ gates over a gate set compatible with Theorem \ref{thm:amplitudetosat} (e.g. Toffoli+H or Clifford+T) or Proposition \ref{thm:algebraicgates}, and let $\ket{\psi}, \ket{\phi}$ be suitable product states. Then $\bra{\psi}U\ket{\phi}$ can be calculated in time sub-exponential in $n$, i.e. asymptotically faster than state-vector simulation, if $G = o(n \sqrt{\log n})$.
\end{corollary}

This bound on $G$ clearly does not contain all
polynomial-sized circuits, but is satisfied by any circuit of depth $o(\sqrt{\log n})$.
Corollary \ref{thm:subexpsim} can thus improve on the tensor-network bound of Markov and Shi, which can be $2^{\Theta(n)}$ even for circuits of depth four \cite[Theorem 1.7]{MS}.

While we phrase the result in terms of calculating amplitudes, the same scaling applies for sampling
and calculating expectation values \cite{Bravyi2022}. However, for a local observable $M$, evaluating $\bra{\psi}U^\dagger MU\ket{\psi}$ requires simulating
only its backward lightcone.
At depth $o(\sqrt{\log n})$, the lightcone contains $o(n)$ qubits and state-vector already gives sub-exponential simulation.

\section{What is the impact on well-known quantum algorithms?}
\label{sec:qualgo}

We will evaluate the impact on quantum advantage for a particular task using the comparison against state-vector simulation given in Corollary \ref{thm:subexpsim}. This is informative because, if Corollary \ref{thm:subexpsim} has worse run-time than state-vector simulation for the parameters of the task in question, then we know it cannot \emph{change} the (asymptotically) best-known classical algorithm for that task, and hence potential for quantum advantage, regardless of what the algorithm is. Thus, in many cases we can avoid having to determine the best-known classical algorithm, which can be difficult. 

The bound $G = o(n\sqrt{\log n})$ is extremely restrictive, and is not satisfied by many quantum algorithms designed for problems currently solved classically. Usually, we have $G = \Omega(n^c)$ with $c > 1$, which implies that state-vector simulation will always asymptotically beat Corollary \ref{thm:subexpsim}. We give the following representative examples:

\begin{enumerate}[label=\roman*)]
    \item For Shor's algorithm $G = \Theta(n^2 \log n)$ using the $O(n \log n)$ multiplication algorithm of \cite{HvdH}. More practical implementations scale as $G = \Theta(n^3)$ \cite{Gidney2025}. Regev's factoring algorithm \cite{Regev2025} and its space-efficient variants can reduce the gate count to $G = \Theta(n^{3/2})$. Note that while Regev's algorithm requires $O(\sqrt{n})$ shots, this only contributes a polynomial factor to the runtime, which can be absorbed into the exponent of Theorem \ref{thm:slopstatement} and hence does not affect the asymptotics here.
    \item For Hamiltonian simulation problems, the complexity can vary wildly. The best scalings we are aware of for electronic structure Hamiltonians are given by \cite{Gunther2026}. For 3D Hamiltonians, their construction has $G = \Omega(n^{5/3}t)$, where $t$ is the simulation time. Even in the simplest 1D case, they find $G = \Omega(n \log n \cdot t)$, and so $G = \Omega(n\sqrt{\log n})$ whenever $t = \Omega(1/\sqrt{\log n})$. Typically we have $t = \Omega(1)$ set by the parameters of the Hamiltonian.
    \item For quantum linear systems with condition number $\kappa$, sparsity $s$, and error $\epsilon$, the query complexity is $\Theta(\kappa \sqrt{s}\log\frac{1}{\epsilon})$ \cite{Mori_2026, patel2026optimaljointlowerbound, BravoPrieto2026}, which is also a lower bound on the number of gates $G$. The parameters vary depending on the application; for example, when used for solving linear partial differential equations, the resulting algorithm scales like $G = \Omega(n^c)$, where, e.g. $c \geq 3$ in \cite[Theorem 1]{Childs2021}.
\end{enumerate}

\noindent On the other hand, there are a few quantum algorithms which may be significantly affected by the new model counting results. For example:

\begin{enumerate}[label=\alph*)]
    \item Algorithms for solving constraint satisfaction problems (CSP), such as 3-SAT, using quantum walks \cite{Montanaro2018} are affected because the classical baselines they are compared against have changed. The method of \cite{Montanaro2018} can accelerate any classical backtracking algorithm for CSPs, providing a Grover-like quadratic speedup. 
    Applied to the state-of-the-art classical methods based on the DPLL algorithm (see e.g \cite{Liu2018}), this had the best known (asymptotic) runtime for 3-SAT prior to the release of \cite{OpenAI,Anthropic}.
    Despite the speedup, the runtime is still exponential in the number of variables, and so this method can no longer provide a quantum advantage, asymptotically.
    \item Sampling measurement outcomes of random low-depth quantum circuits has been used as an experimental demonstration of a task that cannot be replicated classically within a reasonable amount of time \cite{DeCross_2025,Sedrakyan2026,Morvan_2024}. The asymptotic regime in which this is known to be classically hard requires circuits of depth $O(\log n)$ \cite{Bremner_2017,Kondo2021}, and it is an open question whether this extends to the $o(\sqrt{\log n})$-depth regime, which would benefit from Corollary \ref{thm:subexpsim}. In either case, a classical simulation can never be asymptotically better in runtime than sampling from the original circuit.
\end{enumerate}

\noindent In summary, the sub-exponential scaling $T_{\#\mathrm{3-SAT}}(v) = \exp \left[O\left({v}/{\sqrt{\log v}}\right)\right]$ is too weak to give asymptotic speedups over state-vector methods for many of the popular problems studied in quantum computing. 

Even if the scaling were vastly improved, say $T_{\#\mathrm{3-SAT}}(v) = \exp \left[O\left(v^c\right)\right]$ for some constant $c \ll 1$, it is still not obvious that simulating the quantum circuit using Theorem \ref{thm:amplitudetosat} would be beneficial. Consider, for instance, factoring: since the corresponding decision problem is in NP, a factoring instance can be encoded directly as a k-SAT instance with $v = O(n \log n)$ variables using the multiplication algorithm of \cite{HvdH}. Solving the k-SAT instance directly in time $T_{\#\mathrm{3-SAT}}(v)$ is asymptotically more efficient than simulating the $O(n^2 \log n)$ or $O(n^{3/2})$-sized quantum circuits of Shor's or Regev's algorithms. Clearly, unless $c \leq \frac{1}{3}$, solving the k-SAT instance directly is still suboptimal, since the general number field sieve has a much better running time of $\exp[O(n^{1/3}(\log n)^{2/3})]$ \cite{Lee2018}. 

\section{{Further questions and limitations}}
\label{sec:discussion}

{Finally, we discuss further implications and limitations of the recent claims.}

\paragraph*{Could simulation be sub-exponential in the T-count?} Circuits containing few non-Clifford gates can be simulated using stabilizer decomposition methods \cite{Bravyi2019} in time $\exp(O(t))$, where $t$ is the T-count. Therefore, it is natural to ask if Theorem \ref{thm:slopstatement} could lead to an algorithm that is sub-exponential in T-count. Unfortunately, this seems unlikely - while the amplitude of any circuit with $t$ T-gates can be written as the overlap of a product state with a $t$-qubit stabilizer state \cite[Section~2.3.1]{Bravyi2019}, an arbitrary stabilizer state requires $G = \Omega(t^2)$ gates to synthesize. This exceeds the $G = o(t \sqrt{\log t})$ requirement in Corollary \ref{thm:subexpsim}, and hence does not lead to a sub-exponential time algorithm.

\paragraph*{Would an algorithm for k-SAT suffice?}

The headline result of \cite{OpenAI,Anthropic} is that the ETH is false. Strictly speaking, proving this requires only a family of algorithms that solve 3-SAT in time $O(c^n)$ for all $c > 1$, and not a sub-exponential time algorithm for \#3-SAT, nor one for 3-SAT. However, if only a sub-exponential algorithm for k-SAT had been obtained, and not \#k-SAT, would this still have been sufficient for simulating quantum circuits? 

Probably not -- Stockmeyer's theorem \cite{Stockmeyer1983} says that \#k-SAT can be solved up to relative error $1 \pm \varepsilon$ in time $O(\poly(v)\varepsilon^{-2} \cdot T_{\mathrm{k-SAT}}(v))$, where $T_{\mathrm{k-SAT}}(v)$ is the time required to solve k-SAT, by a probabilistic algorithm which makes $O(\poly(v)\varepsilon^{-2})$ calls to a k-SAT solver. This initially seems promising, but unfortunately applying it to the approach in Theorem \ref{thm:amplitudetosat} fails in Lemma \ref{thm:nonbinarytobinary}: a relative-error estimate of each term in the sum does not guarantee an additive-error estimate of the sum, as large cancellations between terms may occur. Indeed, if this were possible, it would imply that $BQP \subseteq BPP^{NP}$, which is unlikely since it is suspected that $BQP$ may lie outside the polynomial hierarchy \cite{Aaronson2009}.

\paragraph*{Would even faster \#k-SAT algorithms affect quantum complexity classes?} As discussed in Section \ref{sec:simulable}, this form of sub-exponential runtime is not closed under polynomial reductions, and its effects are hence limited to fine-grained complexity. In order for it to be closed (in the sense that a polynomial reduction modifies only the base of the exponent), a quasi-polynomial algorithm for \#k-SAT would be required, i.e. running in time $\exp(O(\poly\log(v)))$. This would show that any quantum advantage can be at most quasi-polynomial (i.e. $BQP \subseteq QP$) but also have further-reaching implications, such as the collapse of the exponential hierarchy \cite{Buhrman1992}. 

\paragraph*{Can the sub-exponential simulation result extend to arbitrary gate sets?}

The method in Proposition \ref{thm:algebraicgates} does not scale favorably for circuits containing many distinct gates (e.g $|\mathcal{B}| = \Omega(G)$, rather than $|\mathcal{B}| = O(1)$), for example Trotterized Hamiltonian simulation. On the other hand, as noted in Section \ref{sec:reduction}, approximating $K$ non-Clifford gates to total precision $\varepsilon$ using Clifford+T gates incurs a non-linear overhead of $O(K \log (K / \varepsilon))$. It follows that when the synthesized circuit has $K\log{\frac{K}{\epsilon}} + G = o(n\sqrt{\log{n}})$ gates, the new model counting results have sub-exponential runtime in $n$. However, in this regime $K = o(n/\sqrt{\log n})$, and stabilizer decomposition methods \cite{Bravyi2019} will be more efficient than both state-vector simulation and Corollary \ref{thm:subexpsim}. Therefore, it remains open whether arbitrary gate sets can be handled efficiently for any constant $\varepsilon$. On the other hand, if $\varepsilon$ scales inverse-polynomially with $G$ (i.e. small, but not arbitrarily small), recent work by Zhang et al. \cite{Zhang2026} shows how this can be done with only constant factor overheads.

\paragraph*{Could these new methods be made practical?} We emphasize that we have only discussed asymptotic runtime so far. Perhaps this is appropriate, since many algorithms achieving the best asymptotics can be extremely slow in practice -- \cite{Dias2026} estimates that for random 3-SAT instances near the satisfiability threshold, the algorithms of \cite{OpenAI,Anthropic} are faster than the $2^n$ cost of naive brute-force only when $n > 2^{5.5 \cdot 10^{11}}$. Additionally, backtracking methods for SAT problems are often much faster than their worst-case asymptotic bounds would suggest \cite{Ganesh2021} -- finite-size performance depends on the encoding and implementation, and requires benchmarking against specialized classical algorithms.

However, it is important to consider whether the algorithms that follow from  OpenAI and Anthropic's results could be improved, to make them competitive in practice. For example, compared to state-vector methods, they use exponentially less memory as the number of qubits increases, even when state-vector methods are asymptotically better in terms of runtime. Additionally, their methods are similar in spirit to Lemma \ref{thm:nonbinarytobinary}, since they encode the original \#3-SAT instance into a tensor network, which is split into many other structurally simpler tensor networks (although with a much more complex post-processing). In particular, \cite[Theorem D.1]{OpenAI} applies this idea to tensor networks with entries in a certain family of polynomial rings. 
If these were shown to encode quantum circuits more efficiently, combining the splitting method of \cite{OpenAI,Anthropic} with state-of-the-art contraction techniques (e.g. \cite{Gray2021,Gray2024}) might render the approach viable in practice.

\paragraph*{Do heuristic methods change the conclusion in Section \ref{sec:qualgo}?} For CSPs, since state-of-the-art classical SAT solvers are highly optimized, and because operations (both physical and logical) on quantum computers are much slower than their classical counterparts, quadratic speedups are generally regarded as not practical \cite{Babbush2021}. Hence, it is unlikely that the quantum walk method of \cite{Montanaro2018} would have been used in practice, regardless.

For random circuit sampling, the results of \cite{OpenAI,Anthropic} may in principle make this type of quantum advantage harder to obtain for very shallow circuits. However, this task does not aim to solve any particular classical problem, and so future demonstrations can choose instead to use deeper circuits enabled by the transition to fault-tolerant quantum computation. Furthermore, this task is usually considered at a fixed size \cite{Dalzell_2020}, so the practical performance of Theorem \ref{thm:slopstatement} becomes important.

\begingroup
\setlength{\bibitemsep}{1.5pt}
\setlength{\emergencystretch}{2em}
\printbibliography
\endgroup

\clearpage
\appendix

\section{Reduction of binary-valued tensor networks to \#3-SAT}
\label{app:binarytosat}

\begin{restatedbinarytosat}
    \binarytosatstatement
\end{restatedbinarytosat}

\begin{proof}[Proof of Lemma~\ref{thm:binarytosat}]
    {Let $D$ be the constant upper bounding each index dimension. Encode each index of dimension $D'\leq D$ by $\lceil\log_2 D'\rceil$ bits,
    using the same bits at both ends of the bond. Set tensor entries to zero for unused bit strings.
    Each original index assignment has exactly one valid binary encoding, so this preserves the
    contraction. Each encoded tensor depends on at most $b=2k\lceil\log_2 D\rceil=O(1)$ bits. Write
    $\mathbf{x}$ for all these bits and $\mathbf{x}_i$ for those incident to $T_i$.}

    Let ${T_i(\mathbf{x}_i)} \in \{0, 1\}$ be the elements of the $i$th
    tensor. We can construct a CNF formula $\phi_i$ such that
    ${\phi_i(\mathbf{x}_i) = T_i(\mathbf{x}_i)}$ with
    $O(2^{{b}}) = O(1)$ clauses, by viewing $T_i$ as a truth table. The
    overall evaluation of $T$ for a fixed setting of the indices is given by
    ${\prod_{i = 1}^t T_i(\mathbf{x}_i) =
    \bigwedge_{i=1}^t\phi_i(\mathbf{x}_i) = \phi(\mathbf{x})}$, and so counting the satisfying
    solutions to $\phi({\mathbf{x}})$ is exactly the same as summing over
    the evaluation of $T$ for all index values, i.e. contracting $T$. Applying the standard Tseytin
    encoding {with equivalences} converts $\phi$ to a sparse 3-CNF formula by
    introducing $O(t\cdot 2^{{b}}) = O(t)$ auxiliary variables, in a way
    that preserves the number of satisfying assignments. 
\end{proof}

\section{Reduction for Clifford+T, Toffoli+H, and related gate sets}
\label{app:nonbinarytobinary}

\begin{restatednonbinarytobinary}
    \nonbinarytobinarystatement
\end{restatednonbinarytobinary}

\begin{proof}[Proof of Lemma~\ref{thm:nonbinarytobinary}]
    {
    In a binary-valued network, every nonzero product of entries is $1$, so
    contraction counts the index assignments giving nonzero products. Here we also choose one
    summand from each selected entry. After taking out the factors $\gamma_i$, every nonzero product
    of chosen summands is a power of $\zeta_d$. For each $r\in\{0,\ldots,2^d-1\}$, we construct a
    binary-valued network $T^{(r)}$ whose contraction counts the choices of original indices and
    summands giving product value $\zeta_d^r$.

    In the entry expansion of Lemma~\ref{thm:nonbinarytobinary}, $j_i$ records the exponent of the summand chosen at $T_i$. We
    introduce the running sums $q_i=(j_1+\cdots+j_i)\bmod 2^d$ to check the total using tensors of
    bounded degree. Each $j_i$ and $q_i$ is a single index taking values in $\{0,\ldots,2^d-1\}$.
    Replace each $T_i$ by two binary-valued tensors:

    \noindent\begin{minipage}[c]{0.40\linewidth}
        \[
        S_i(\mathbf{x}_i,j_i)=\beta^{(i)}_{j_i,\mathbf{x}_i},
        \]
    \end{minipage}\hfill
    \begin{minipage}[c]{0.57\linewidth}
        \[
        P_i(j_i,q_{i-1},q_i)=
        \begin{cases}
            1,&q_i\equiv q_{i-1}+j_i\pmod{2^d},\\
            0,&\text{otherwise}.
        \end{cases}
        \]
    \end{minipage}
    \begin{center}
        \begin{minipage}{\linewidth}
            \centering
            \input{figures/lemma3-counting-network-two-panel.tex}
        \end{minipage}
    \end{center}
    We obtain $T^{(r)}$ by retaining the original index connections and adding two tensors that
    enforce $q_0=0$ and $q_t=r$, so that $r=(j_1+\cdots+j_t)\bmod 2^d$.

    Since the $q_i$ are uniquely determined by the $j_i$, each valid choice of original indices and
    summands is counted once. Thus $\mathrm{Contract}(T^{(r)})$ counts the terms with value
    $\zeta_d^r$. Multiplying this count by $\zeta_d^r$ gives their total contribution. Summing over
    $r$ and restoring the scalar factors gives
    \[
        \mathrm{Contract}(T)
        =\left(\prod_{i=1}^{t}\gamma_i\right)
        \sum_{r=0}^{2^d-1}\zeta_d^r \cdot \mathrm{Contract}\!\left(T^{(r)}\right).
    \]
    Each network contains $2t+2$ tensors, of degree at most $2k+1$ for $S_i$, three for $P_i$, and
    one for the boundary tensors. The bond dimensions are $2$ or $2^d$. Since $k$ and $d$ are fixed,
    each network can be constructed in $O(t)$ time and satisfies Lemma \ref{thm:binarytosat}.}
\end{proof}

\section{{Reduction for arbitrary gate sets}}
\label{app:algebraicgates}

{

\begin{restatedalgebraicgates}
    \algebraicgatesstatement
\end{restatedalgebraicgates}
}

{
\begin{proof}[Proof of Proposition~\ref{thm:algebraicgates}]
    As in Lemma \ref{thm:nonbinarytobinary}, we recover the contraction by counting terms with a
    common product value. Let $a_1,\ldots,a_s$ be the possible nonzero entries and $m_j$ the number
    of occurrences of $a_j$ in a nonzero term. The term then has value $a_1^{m_1}\cdots a_s^{m_s}$,
    so we count terms by recording these multiplicities \cite[Section 2.2]{CaiChen2011}.

    We first construct the tensor network $T$ according to Lemma \ref{thm:circuittononbinary}. It
    has $N=O(n+G)$ bounded-degree tensors $T_1,\ldots,T_N$, with nonzero entries in this fixed set.
    Each assignment of the original indices selects one entry from each tensor. If it selects no
    zeros, then $\sum_j m_j=N$. For each such vector $\boldsymbol{m}=(m_1,\ldots,m_s)$, we construct
    a formula $F_{\boldsymbol{m}}$ whose number of satisfying assignments, denoted
    $\#F_{\boldsymbol{m}}$, counts the index assignments selecting exactly $m_j$ entries equal to
    $a_j$, for every $j$. Thus
    \[
        \mathrm{Contract}(T)
        =\sum_{\substack{\boldsymbol{m}\in\mathbb{Z}_{\geq 0}^{s}\\m_1+\cdots+m_s=N}}
        \#F_{\boldsymbol{m}}\prod_{j=1}^{s}a_j^{m_j}.
    \]
    There are $\binom{N+s-1}{s-1}=O(N^{s-1})$ possible vectors $\boldsymbol{m}$, so the number of
    queries is polynomial for fixed $s$.

    To construct $F_{\boldsymbol{m}}$, let $\mathbf{x}_i$ denote the indices of $T_i$ and introduce
    binary indices $b_{i,1},\ldots,b_{i,s}$, where $b_{i,j}$ indicates whether the entry selected
    from $T_i$ is $a_j$. Replace $T_i$ by a Boolean-valued tensor
    $L_i(\mathbf{x}_i,b_{i,1},\ldots,b_{i,s})$. Its entry is $1$ exactly when
    $T_i(\mathbf{x}_i)\ne0$ and, for every $j$, $b_{i,j}=1$ if and only if $T_i(\mathbf{x}_i)=a_j$.
    Each $L_i$ has constant size. Require
    \[
        \sum_{i=1}^{N}b_{i,j}=m_j,\qquad j=1,\ldots,s.
    \]
    As in Lemma~\ref{thm:nonbinarytobinary}, we use a \emph{counting gadget} to check each sum.
    Fixing all totals to $\boldsymbol{m}$ makes the whole network contract to $\#F_{\boldsymbol{m}}$.
    \begin{center}
        \begin{minipage}{\linewidth}
            \centering
            \input{figures/proposition5-entry-counter.tex}
        \end{minipage}
    \end{center}
    We build each counting gadget as a balanced tree by adding the input bits in pairs, then
    repeatedly adding pairs of partial sums until one total remains. An unpaired value passes to
    the next level unchanged. The first additions form level $\ell=1$. The diagram below shows
    one counting gadget: each wire carries a partial sum, and each $+$ box adds its two inputs.
    \begin{center}
        \begin{minipage}{\linewidth}
            \centering
            \input{figures/proposition5-counting-gadget.tex}
        \end{minipage}
    \end{center}
    Each adder at level $\ell$ adds $O(\ell)$-bit numbers and can be built from $O(\ell)$
    Boolean-valued tensors. There are $O(N/2^\ell)$ additions at that level, so
    each counting gadget contains $\sum_{\ell\geq 1}O(N\ell/2^\ell)=O(N)$ tensors of bounded degree
    with binary indices. Since $s$ is fixed, the whole network contains
    $O(N)$ tensors. The original indices uniquely determine all added indices, so each accepted
    assignment is counted once. Applying Lemma~\ref{thm:binarytosat} gives $F_{\boldsymbol{m}}$
    with $O(N)$ variables and clauses. {The final weighted sum can be evaluated
    using polynomially many arithmetic operations, assuming exact arithmetic on the fixed entries.}
\end{proof}
}

\end{document}

%% file: figures/lemma3-counting-network-two-panel.tex
% Lemma 3: the full network and the expansion of its counting gadget.
\begin{tikzpicture}[
    x=1cm,y=1cm,font=\small,
    tensor/.style={draw=black,line width=0.7pt,rounded corners=1.5pt,
        fill=black!5,minimum width=0.85cm,minimum height=0.52cm,
        inner sep=3pt,text=black},
    boundary/.style={tensor,fill=white,minimum width=0.65cm},
    gadget/.style={draw=black,dashed,line width=0.7pt,rounded corners=3pt,
        fill=black!2,minimum width=2.65cm,minimum height=0.90cm,
        inner sep=4pt,text=black,align=center},
    bond/.style={draw=black,line width=0.7pt},
    index/.style={font=\small,text=black,inner sep=2pt},
    panel/.style={font=\small,text=black,anchor=base,inner sep=0pt}
]
    \node[panel] at (3.20,3.10) {(a) The network $T^{(r)}$};

    % The original network connects to the gadget through the summand labels.
    \draw[draw=black,dashed,rounded corners=3pt,line width=0.65pt]
        (0.25,1.98) rectangle (6.15,2.57);
    \node[text=black] at (3.20,2.275)
        {Original index connections};

    \foreach \i/\x in {1/0.90,2/3.00,t/5.50}{
        \node[tensor] (S\i) at (\x,1.32) {$S_{\i}$};
        \foreach \offset in {-0.13,0,0.13}{
            \draw[bond] (\x+\offset,1.58) -- (\x+\offset,1.98);
        }
        \node[index,anchor=east] at (\x-0.22,1.78) {$\mathbf{x}_{\i}$};
    }
    \node[text=black] at (4.25,1.32) {$\cdots$};

    \node[gadget] (G) at (3.20,-0.45) {modular\\counting gadget};
    \foreach \i/\x/\port/\offset in {1/0.90/2.40/-0.80,2/3.00/3.20/0,t/5.50/4.00/0.80}{
        \draw[bond] (S\i.south)
            .. controls (\x,0.65) and (\port,0.75)
            .. (\port,0.45) -- ([xshift=\offset cm]G.north);
        \node[index,anchor=west] at (\port+0.04,0.40) {$j_{\i}$};
    }
    \node[boundary] (result) at (3.20,-1.85) {$\bra{r}$};
    \draw[bond] (G.south) -- node[index,right] {$q_t$} (result.north);

    \begin{scope}[xshift=7.15cm]
        \node[panel] at (4.00,3.10) {(b) Expansion of the counting gadget};

        % The initial zero is internal; the j_i inputs and q_t output stay open.
        \draw[draw=black,dashed,rounded corners=3pt,line width=0.7pt,
            fill=black!2] (-0.10,-0.35) rectangle (8.10,1.25);
        \node[boundary] (zero) at (0.45,0.45) {$\ket{0}$};
        \node[tensor] (P1) at (1.95,0.45) {$P_1$};
        \node[tensor] (P2) at (3.95,0.45) {$P_2$};
        \node[text=black] (dots) at (5.40,0.45) {$\cdots$};
        \node[tensor] (Pt) at (6.90,0.45) {$P_t$};

        \foreach \i/\x in {1/1.95,2/3.95,t/6.90}{
            \draw[bond] (\x,1.95) -- (P\i.north);
            \node[index,above] at (\x,1.95) {$j_{\i}$};
        }
        \draw[bond] (zero.east) -- node[index,below] {$q_0$} (P1.west);
        \draw[bond] (P1.east) -- node[index,below] {$q_1$} (P2.west);
        \draw[bond] (P2.east) -- node[index,below] {$q_2$} (dots.west);
        \draw[bond] (dots.east) -- node[index,below] {$q_{t-1}$} (Pt.west);
        \draw[bond,rounded corners=3pt] (Pt.east) -- (7.65,0.45) -- (7.65,-1.20);
        \node[index,right] at (7.65,-0.85) {$q_t$};
    \end{scope}
\end{tikzpicture}

%% file: figures/proposition5-entry-counter.tex
% Proposition 5: entry lookup tensors and their multiplicity counters.
\begin{tikzpicture}[
    x=1cm,y=1cm,font=\small,
    tensor/.style={draw=black,line width=0.7pt,rounded corners=1.5pt,
        fill=black!5,minimum width=0.90cm,minimum height=0.55cm,
        inner sep=3pt,text=black},
    gadget/.style={draw=black,dashed,line width=0.7pt,rounded corners=3pt,
        fill=black!2,minimum width=2.65cm,minimum height=0.90cm,
        inner sep=4pt,text=black,align=center},
    bond/.style={draw=black,line width=0.7pt},
    crossing/.style={bond,preaction={draw=white,line width=2.4pt}},
    index/.style={font=\small,text=black,inner sep=1pt}
]
    \draw[draw=black,dashed,rounded corners=3pt,line width=0.65pt]
        (-0.85,3.10) rectangle (10.85,3.75);
    \node[text=black] at (5,3.425) {Original index connections};

    \foreach \i/\x in {1/0,2/3.8,N/10}{
        \node[tensor] (L\i) at (\x,2.35) {$L_{\i}$};
        \foreach \offset in {-0.13,0,0.13}{
            \draw[bond] (\x+\offset,2.625) -- (\x+\offset,3.10);
        }
        \node[index,anchor=east] at (\x-0.22,2.87) {$\mathbf{x}_{\i}$};
    }
    \node[text=black] at (6.90,2.35) {$\cdots$};

    % Each dashed region abbreviates an entire Boolean counting subnetwork.
    \foreach \j/\x in {1/0,2/3.8,s/10}{
        \node[gadget] (C\j) at (\x,-0.70) {counting gadget\\for $a_{\j}$};
        \node[tensor,fill=white] (M\j) at (\x,-1.85) {$\bra{m_{\j}}$};
        \draw[bond] (C\j.south) -- (M\j.north);
    }
    \node[text=black] at (6.90,-0.70) {$\cdots$};

    % White under-strokes distinguish crossing bonds from junctions.
    \foreach \j/\cx/\sourceoffset in {1/0/-0.24,2/3.8/0,s/10/0.24}{
        \foreach \i/\lx/\targetoffset in {1/0/-0.80,2/3.8/0,N/10/0.80}{
            \draw[crossing]
                (\lx+\sourceoffset,2.075)
                .. controls (\lx+\sourceoffset,1.30) and (\cx+\targetoffset,1.35)
                .. (\cx+\targetoffset,0.55)
                -- ([xshift=\targetoffset cm]C\j.north);
        }
    }
    \foreach \j/\cx in {1/0,2/3.8,s/10}{
        \foreach \i/\offset in {1/-0.80,2/0,N/0.80}{
            \node[index,fill=white] at (\cx+\offset,0.22) {$b_{\i,\j}$};
        }
    }
\end{tikzpicture}

%% file: figures/proposition5-counting-gadget.tex
% Proposition 5: expand one counting gadget, omitting repeated branches and levels.
% Each wire carries one partial-sum index; each + box enforces addition.
% The proof expands these indices and adders into a network with binary indices.
\begin{tikzpicture}[
    x=1cm,y=1cm,font=\small,
    adder/.style={draw=black,line width=0.7pt,rounded corners=1.5pt,
        fill=black!5,minimum width=0.70cm,minimum height=0.45cm,
        inner sep=3pt,text=black},
    boundary/.style={adder,fill=white,minimum width=0.90cm,minimum height=0.55cm},
    bond/.style={draw=black,line width=0.7pt},
    index/.style={font=\small,text=black,inner sep=2pt}
]
    \draw[draw=black,dashed,rounded corners=3pt,line width=0.7pt,
        fill=black!2] (-0.40,-1.75) rectangle (9.25,1.65);

    % Level 1 adds pairs of input bits.
    \foreach \i/\x in {1/0.575,2/2.875,3/5.975,4/8.275}{
        \node[adder] (A\i) at (\x,1.10) {$+$};
    }
    \foreach \i/\x/\target/\port in {
        1/0/1/-0.20,2/1.15/1/0.20,
        3/2.30/2/-0.20,4/3.45/2/0.20,
        {N-3}/5.40/3/-0.20,{N-2}/6.55/3/0.20,
        {N-1}/7.70/4/-0.20,N/8.85/4/0.20}{
        \node[index,above] at (\x,1.95) {$b_{\i,j}$};
        \draw[bond] (\x,1.93) -- ([xshift=\port cm]A\target.north);
    }
    \node[index,above] at (4.425,1.95) {$\cdots$};
    \node[index] at (4.425,1.10) {$\cdots$};

    % Level 2 adds the partial sums in pairs.
    \node[adder] (B1) at (1.725,0) {$+$};
    \node[adder] (B2) at (7.125,0) {$+$};
    \foreach \i/\target/\port in {1/1/-0.20,2/1/0.20,3/2/-0.20,4/2/0.20}{
        \draw[bond] (A\i.south) -- ([xshift=\port cm]B\target.north);
    }
    \node[index] at (4.425,0) {$\cdots$};

    % Ellipses on the branches stand for the remaining levels of addition.
    \node[adder] (total) at (4.425,-1.30) {$+$};
    \draw[bond] (B1.south) -- node[index,midway,sloped,fill=black!2,
        inner xsep=3pt,inner ysep=0pt] {$\cdots$} ([xshift=-0.20cm]total.north);
    \draw[bond] (B2.south) -- node[index,midway,sloped,fill=black!2,
        inner xsep=3pt,inner ysep=0pt] {$\cdots$} ([xshift=0.20cm]total.north);
    \node[boundary] (selected) at (4.425,-2.40) {$\bra{m_j}$};
    \draw[bond] (total.south) -- (selected.north);

    \foreach \level/\y in {1/1.10,2/0}{
        \node[index,anchor=west] at (9.55,\y) {$\ell=\level$};
    }
    \node[index] at (9.90,-0.65) {$\vdots$};
    \node[index,anchor=west] at (9.55,-1.30) {$\ell=\lceil\log_2 N\rceil$};
\end{tikzpicture}